\documentclass[11pt,a4paper]{article}
\pdfoutput=1
\usepackage{jheppub}
\usepackage{graphicx} 
\usepackage{epstopdf} 
\usepackage{dcolumn} 
\usepackage{xcolor} 
\usepackage{amsmath,amssymb} 
\usepackage{hyperref} 
\usepackage[utf8]{inputenc} 
\usepackage[english]{babel} 
\usepackage[displaymath, mathlines]{lineno} 
\usepackage{blindtext} 
\usepackage{ifthen} 
\usepackage{orcidlink} 
\usepackage{adjustbox}
\graphicspath{{figures/}} 
\usepackage{lmodern}
\newboolean{articletitles}
\setboolean{articletitles}{true} 

\input{belle2-symbols}

\begin{document}

\vspace*{-3\baselineskip}
\resizebox{!}{2cm}{\includegraphics{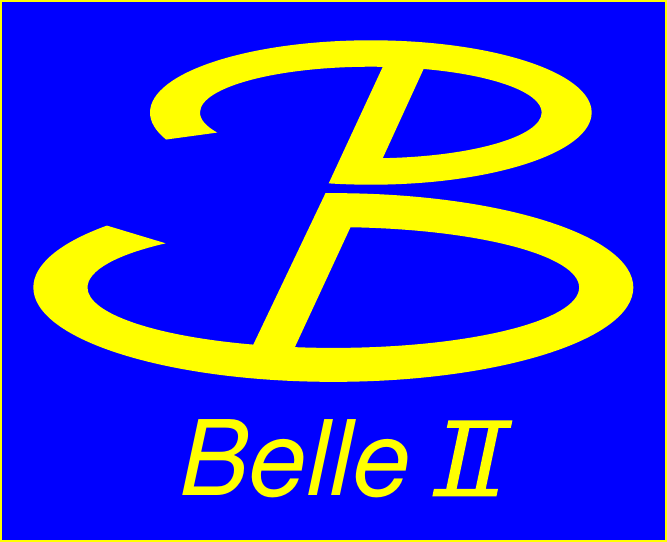}}
\title{Measurement of time-dependent \CP asymmetries in $\Bz \to \KS \: \pip \pim \gamma$ decays at Belle and Belle~II}
\collaboration{The Belle and Belle II Collaborations}
  \author{M.~Abumusabh\,\orcidlink{0009-0004-1031-5425},} 
  \author{I.~Adachi\,\orcidlink{0000-0003-2287-0173},} 
  \author{L.~Aggarwal\,\orcidlink{0000-0002-0909-7537},} 
  \author{H.~Ahmed\,\orcidlink{0000-0003-3976-7498},} 
  \author{Y.~Ahn\,\orcidlink{0000-0001-6820-0576},} 
  \author{H.~Aihara\,\orcidlink{0000-0002-1907-5964},} 
  \author{N.~Akopov\,\orcidlink{0000-0002-4425-2096},} 
  \author{S.~Alghamdi\,\orcidlink{0000-0001-7609-112X},} 
  \author{M.~Alhakami\,\orcidlink{0000-0002-2234-8628},} 
  \author{K.~Amos\,\orcidlink{0000-0003-1757-5620},} 
% ...

\abstract{\unboldmath
We present a measurement of the time-dependent \CP asymmetry in $\Bz \to \KS \: \pip \pim \gamma$ decays using a data set of $365~\invfb$ recorded by the Belle~II experiment and the final data set of $711~\invfb$ recorded by the Belle experiment at the \ups resonance. The direct and mixing-induced time-dependent \CP violation parameters $C$ and $S$ are determined along with two additional quantities, \Sp and \Sm, defined in the two halves of the $m^2(\KS \pip)-m^2(\KS \pim)$ plane. The measured values are $C = -0.17 \pm 0.09 \pm 0.04$, $S = -0.29 \pm 0.11 \pm 0.05$, $\Sp = -0.57 \pm 0.23 \pm 0.10$ and $\Sm = 0.31 \pm 0.24 \pm 0.05$, where the first uncertainty is statistical and the second systematic.

\clearpage
}

\maketitle
\flushbottom

\section{Introduction}
\label{intro}
Flavor-changing neutral currents, in particular ${b \to s \gamma}$ transitions, are sensitive probes of the Standard Model (SM) of particle physics~\cite{Atwood:1997zr,Kou:2013gna,Haba:2015gwa}. The emitted photon in these transitions is predominantly left-handed. A right-handed photon is predicted to occur only due to a chirality flip of the outgoing $s$-quark line, which is suppressed by a factor proportional to $m^2_s/m^2_b$. Therefore, in decays of \Bzb and \Bz mesons to a \CP eigenstate and a photon, denoted as $\Bz\to f_{\CP}\gamma$, the mixing-induced \CP violation is predicted to be very small. Physics beyond the SM may enhance the mixing-induced \CP violation by increasing the right-handed current contribution to these transitions. Measurement of the mixing-induced \CP asymmetry in these decay channels probes those non-SM processes that result in larger \CP asymmetries. 

At the KEKB and SuperKEKB colliders, pairs of $B$~mesons in a coherent quantum state are produced through $\ep\en$ collisions at the \ups resonance. The $B$~mesons are referred to as $B_{\mathrm{sig}}$ and $B_{\mathrm{tag}}$, where $B_{\mathrm{sig}}$ is the meson of interest and $B_{\mathrm{tag}}$ is the meson whose information is used to infer the flavor ($\Bz$ or $\Bzb$) of 
$B_{\mathrm{sig}}$ at the time the $B_{\mathrm{tag}}$ decays.
The time-dependent \CP asymmetry in neutral \B mesons decaying to a final state $f_{\CP}\gamma$ is defined as

\begin{equation*}
    \mathcal{A}_{CP}(\Delta t) =
    \frac{\Gamma(B_{\mathrm{tag}=\Bz}(\Delta t) \rightarrow f_{CP}\gamma) -
        \Gamma(B_{\mathrm{tag}=\Bzb}(\Delta t) \rightarrow f_{CP}\gamma)}
        {\Gamma(B_{\mathrm{tag}=\Bz}(\Delta t) \rightarrow f_{CP}\gamma) +
        \Gamma(B_{\mathrm{tag}=\Bzb}(\Delta t) \rightarrow f_{CP}\gamma)}
\end{equation*}
where $\Gamma(B_{\mathrm{tag}=\Bz}(\dt))$ is the decay rate of a \B~meson for which its companion has been tagged as a \Bz at decay, and $\dt \equiv t_{\mathrm{sig}}-t_{\mathrm{tag}}$ corresponds to the difference between the proper decay times of the $B_{\mathrm{sig}}$ and $B_{\mathrm{tag}}$. It can be parameterized as follows:
\begin{equation}
    {\cal A}_{CP}(\dt) = S \sin(\Delta m \dt) - C \cos(\Delta m \dt),
\label{eq:aCP}
\end{equation}
where $S$ and $C $ are known as the mixing-induced and direct \CP violation parameters, respectively, and $\Delta m$ is the mass difference between the heavy and light mass eigenstates of the neutral $B$ mesons. The BaBar and Belle experiments have reported measurements~\cite{BaBar:2015chw,Belle:2008fjm} of the time-dependent \CP asymmetries for the ${\Bz \to \KS \pip \pim \gamma}$ decay, with uncertainties at the level of 25\% (the Belle measurement was performed on a subset of the full Belle data).

This paper presents the combined measurement of the time-dependent \CP asymmetry in $\Bz \to \KS \pip \pim \gamma$ decays using the entire Belle dataset, $711~\invfb$, and $365~\invfb$ of data recorded between 2019 and 2022 by Belle~II.

The channel of interest is the \ $\Bz \to K_{\mathrm{res}}\gamma \to(\KS \pip \pim) \gamma$ decay, where the intermediate resonance $K_{\mathrm{res}}$ decays into a \KS and a charged-pion pair. Among the many possible intermediate resonances only those that decay through the two-body $K_{\mathrm{res}}\to\KS\rho^0$ channel are true \CP eigenstates. For this measurement, we only consider the decay of \KS to two charged pions, $\KS\to\pip\pim$, while the $\rho^0$ meson is reconstructed using its main decay mode, $\rho^0\to\pip\pim$. The time-dependent \CP asymmetry we measure has, in addition to the \CP eigenstate ${\Bz\to\KS\rho^0\gamma}$ mode, contributions from non-\CP eigenstates involving strange meson resonances.  The most important of these are
\begin{align} 
    \Bz\to K_{\mathrm{res}}\gamma &\to (K^{*\pm}\pi^{\mp})\gamma \to ((\KS\pi^{\pm})\pi^\mp)\gamma \label{eq:kstarpath}\\
    \Bz\to K_{\mathrm{res}}\gamma &\to ((K\pi)_0^\pm\pi^{\mp})\gamma \to ((\KS\pi^{\pm})\pi^\mp)\gamma, \label{eq:swavepath}
\end{align}
where $(K\pi)_0^\pm$ represents a $K\pi$ pair in an S-wave configuration.
The contributions of these modes to the effective time-dependent \CP asymmetries could be determined through a full amplitude analysis~\cite{BaBar:2015chw,Belle:2008fjm,Akar:2018zhv} of the isospin partner mode ${B^+\to K^+ \pip \pim \gamma}$, which would provide a full description of the amplitudes and interferences present in the $K_{\mathrm{res}}$ system. No amplitude analysis is performed in this work, but the isospin partner channel is used to validate aspects of the analysis strategy.

In addition to the measurement of the time-dependent \CP asymmetries, $S$ and $C$, we measure two new \CP observables, proposed in Ref.~\cite{Akar:2018zhv}.
We construct these new \CP observables from a combined measurement of the time-dependent \CP asymmetry in two halves of the $({m^2(\KS \pip), m^2(\KS \pim))}$ plane defined by the inequalities ${m^2(\KS \pip) \lessgtr m^2(\KS \pim)}$. To ease the notation, in subsequent sections we refer to the half-plane where ${m^2(\KS \pip) > m^2(\KS \pim)}$ as the ``up'' half while the opposite half is referred to as the ``down" half. Following this notation, the two new observables are defined as
\begin{align*}
    S^+ = S^{\mathrm{up}}+S^{\mathrm{down}} \\
    S^- = S^{\mathrm{up}}-S^{\mathrm{down}},
\end{align*}
where $S^{\mathrm{up}}$ and $S^{\mathrm{down}}$ are the values of $S$, defined in Eq.~\ref{eq:aCP}, measured in the two half-planes. These new \CP observables can be combined with two parameters $(a, b)$, that can be extracted from an amplitude analysis of the isospin partner channel. These parameters describe the different proportions and the interference properties of the \CP mode with respect to the non-\CP-eigenstate modes. These four quantities, taken together, could provide constraints on the photon polarization in the ${\Bz \to K_{\mathrm{res}}\gamma \to(\KS \pip \pim) \gamma}$ mode.

The candidate selection and fit strategy are developed and validated before accessing the data containing the signal mode. Throughout this paper, the inclusion of the charge conjugate decay mode is implied unless otherwise specified.

The outline of this paper is as follows: we present the detectors, Belle and Belle~II, and the corresponding data samples in Sec.~\ref{exps} and Sec.~\ref{datasets}, respectively. The reconstruction and selection of candidates from the decay channel of interest, $\Bz \to \KS \pip \pim \gamma$, and of its isospin partner are described in Sec.~\ref{candidates}. We discuss the strategy to extract the \CP observables and the associated uncertainties in Sec.~\ref{fitstrat}. Finally, we present the results of the measurement and our conclusions in Sec.~\ref{results}.

\section{The Belle and Belle~II experiments}
\label{exps}

The Belle and Belle~II detectors both have a cylindrical geometry whose symmetry axis $z$ is nearly aligned with the electron beam direction at the interaction point.  The polar angle $\theta$ is defined relative to the $z$ axis.  

The Belle detector~\cite{Abashian2002117,Belle:2012iwr} was located at the KEKB $\ep\en$ accelerator~\cite{Abe:2013kxa}, which collided electrons and positrons at and near the $\Upsilon(4S)$ resonance with beam energies of $8$~GeV and $3.5$~GeV, respectively. It recorded data from 1999 to 2010.
The Belle detector was a large-solid-angle magnetic spectrometer composed of a silicon vertex detector (SVD), where two different configurations of the silicon vertex detector and beam pipe radius were used over the course of the experiment, a central drift chamber (CDC), an array of aerogel threshold Cherenkov counters (ACC), a barrel-shaped arrangement of time-of-flight (TOF) scintillation counters, and an electromagnetic calorimeter (ECL) comprised of CsI(Tl) crystals located inside a superconducting solenoid coil that provided a magnetic field of $1.5$~T. An iron flux-return yoke, placed outside the coil, was instrumented with resistive-plate chambers to detect $K^0_L$ mesons and identify muons. The SVD and CDC were used to reconstruct charged particle tracks and vertices, the ACC and TOF, along with ionization energy-loss (${dE}/{dx}$) measurements from the CDC, were used for charged particle identification (PID) purposes, and photons were reconstructed from clusters in the ECL.

The Belle~II detector~\cite{Abe:2010gxa} is located at the SuperKEKB accelerator, which collides electrons and positrons at and near the $\Upsilon(4S)$ resonance~\cite{Akai:2018mbz} with beam energies of 7~GeV and 4~GeV, respectively. The Belle~II detector~\cite{Abe:2010gxa} has a cylindrical geometry and includes a two-layer silicon-pixel detector~(PXD) surrounded by a four-layer double-sided SVD~\cite{Belle-IISVD:2022upf} and a 56-layer CDC. These detectors reconstruct tracks from charged particles.  Only one sixth of the second layer of the PXD was installed for the data analysed here. Surrounding the CDC, which also provides ${dE}/{dx}$ energy-loss measurements, there is a time-of-propagation counter~(TOP)~\cite{Kotchetkov:2018qzw} in the central region and an aerogel-based ring-imaging Cherenkov counter~(ARICH) in the forward region.  These detectors provide charged-particle identification.  Surrounding the TOP and ARICH there is an ECL based on CsI(Tl) crystals that primarily provides energy and timing measurements for photons and electrons. Outside of the ECL there is a superconducting solenoid magnet. Its flux return is instrumented with resistive-plate chambers and plastic scintillator modules to detect muons, $K^0_L$ mesons, and
neutrons. The solenoid magnet provides a 1.5~T magnetic field that is oriented parallel to the $z$ axis.

\section{Datasets}
\label{datasets}

This measurement is based on $365~\invfb$ of data recorded by the Belle~II experiment and $711~\invfb$ recorded by the Belle experiment, both at the \ups resonance. Additional data samples, $43~\invfb$ for Belle~II and $86~\invfb$ for Belle, recorded below the $\ups\to\B\Bbar$ threshold are used for background studies.

We use several Monte-Carlo simulated samples (MC samples) to model, study, and validate different parts of the measurement process. Samples of $\ep\en \to \ups \to \B\Bbar$ are generated, together with the subsequent particle decays, using the \evtgen~\cite{Lange:2001uf} program interfaced to the \pythia8 (\pythia6) software~\cite{Sjostrand:2014zea} for Belle~II (Belle). Quark-antiquark pairs created in the $\ep\en \to q\overline{q}$ process ($q = u,d,s,c$), referred to as continuum events, are generated with the \kkmc~\cite{Jadach:1999vf} package together with \pythia8 (\pythia6) to handle the fragmentation process. The detector response is simulated by the \geant4 (\geant3)~\cite{Agostinelli:2002hh} package. The simulated samples include the effects of beam-induced background, as described in Ref.~\cite{Lewis:2018ayu}. We use MC samples from generic $\ep\en$ collisions,  i.e., combining $\Bz\Bzb$, $\Bp\Bm$ and $q\overline{q}$ samples, and also specific samples of $\B\Bbar$, where one of the $B$ mesons decays into a specified mode of interest (signal MC). We use generic MC samples corresponding to four (six) times the data luminosity for the Belle~II (Belle) analysis. In addition, in Belle, we use a sample of specific rare $\B\Bbar$ decays with fifty times the recorded Belle luminosity to study other decay modes that may affect our measurement as background sources. The signal MC samples are substantially larger than the generic ones, and are generated with a single intermediate resonance, $K_{\mathrm{res}} = K_{1}(1270) \to \KS \rho$ for both Belle and Belle~II analyses.

We use the Belle~II analysis software~\cite{basf2-zenodo,Kuhr:2018lps} to process both collision data and the simulated MC samples in Belle~II, while we use a specific Belle analysis and software framework for Belle.

\section{Candidate selection}
\label{candidates}

The trigger system selects events based on the number of charged and neutral particles, along with the total ECL energy deposition, and retains hadronic $\B\Bbar$ events with an efficiency close to $100\%$ for the signal decay mode.

In each event, the $B_{\mathrm{sig}}$ meson is reconstructed first, and the remaining reconstructed particles are assigned to the $B_{\mathrm{tag}}$ decay. The $B_{\mathrm{sig}}$ candidate selection is optimized to enhance the signal contribution relative to background contributions. While the selection criteria are largely similar between Belle and Belle~II, differences emerge due to differences in detector performance and background conditions. Candidate \ups events are required to have a minimum of 4 (3) charged tracks for Belle~II (Belle) and a visible energy of at least 4 \gev.

Candidate signal photons are reconstructed from ECL clusters with no associated track and with energies higher than $1.5\;(1.4)\gev$ for Belle~II (Belle). An upper bound is set on the energy of these ECL clusters at $4\;(3.5)\gev$ to remove photon candidates arising from beam background. The photon polar angle must satisfy $\cos{\theta_{\gamma}} \in [-0.87, 0.95]$ $\left([-0.65, 0.86]\right)$. Multivariate classifiers that combine information from the photon candidate and the rest of the event are used to remove photon candidates from $\piz\to\gamma\gamma$ and $\eta\to\gamma\gamma$ decays~\cite{Belle:2004stc,Belle:2024qhe}.

The pion candidates that arise directly from the $K_{\mathrm{res}}$ decay, and not through the subsequent \KS decay, are referred to as prompt pions.  The corresponding charged tracks are required to point to the beam interaction region with a longitudinal distance smaller than $3$~cm and a transverse distance smaller than $0.5 \; (1)$ cm for Belle~II (Belle) with respect to the point of closest approach of the track to the beam interaction region. Using the PID system, a loose requirement is placed on these tracks to be compatible with a pion hypothesis. 

The \KS candidates are formed by combining two opposite-sign charged particle tracks displaced with respect to the IP, and their invariant mass, assuming they are pions, is required to be within $\pm20\mevcc$ of the known \KS mass. The \KS invariant mass resolution is about $6\mevcc$. Additionally, a multivariate classifier is used to determine the likelihood of the candidate to be a \KS, using the kinematic information from the tracks and their combination, the flight length of the \KS candidate and the number of hits in the vertex detectors. In Belle the algorithm is based on a NeuroBayes Neural Network~\cite{Feindt:2006pm,Belle:2018xst}, while in Belle~II we use a Boosted Decision Tree (BDT)~\cite{Belle-II:2024uzp}.

The $K_{\mathrm{res}}$ candidates are constructed by summing the four momenta of the prompt pions and \KS candidates. The $K_{\mathrm{res}}$ candidates are required to have an invariant mass ${m_{K_{\mathrm{res}}}\in [0.9, 1.8] \gevcc}$, which allows higher mass structures arising from $B$ decay backgrounds containing a charm meson to be removed. Their momentum in the center of mass (c.m.)\ frame is required to satisfy $p^*_{K_{\mathrm{res}}} \in [1, 3.5]~\gevc$; no candidates arising from the signal mode are expected outside this range.
The $B_{\mathrm{sig}}$ candidates are constructed by combining the four-momenta of the $K_{\mathrm{res}}$ and photon candidates. The reconstructed decay vertex of the $B_{\mathrm{sig}}$ candidate is indistinguishable from the decay vertex of the $K_{\mathrm{res}}$, since the latter decays via the strong force. We determine the $B_{\mathrm{sig}}$ vertex position by performing a fit, using only the prompt pion tracks, while constraining the $B_{\mathrm{sig}}$ to come from the beam interaction region. We only keep $B_{\mathrm{sig}}$ candidates for which the vertex fit has converged. The effect of adding the $\KS$ information into the reconstructed vertex is found to be negligible, so this information was not used. One advantage of this choice is that the ${B^+\to K^+ \pip \pim \gamma}$ decay can be used as a control mode, since the same reconstruction can be obtained as in the signal mode vertex by using only the two pions. The momenta of all particles, including the photon, in the $B_{\mathrm{sig}}$ decay are recomputed after the vertex fit. Additionally, the $B_{\mathrm{sig}}$ are required to have $\mbc >5.20~\gevcc$ and $\Delta E \in [-0.2, 0.2]~\gev$, where the beam-energy-constrained mass and energy difference are defined as $\mbc \equiv \sqrt{(\sqrt{s}/2)^2 - p^{*2}_{B}}$ and $\Delta E \equiv E^*_{B} - \sqrt{s}/2$. The variables $p^*_{B}$, $E^*_{B}$ are the momentum and the energy of the \B meson in the c.m.\ frame, respectively, and $\sqrt{s}$ is the c.m.\ energy of the $\ep\en$ collision. An additional requirement is imposed on the invariant mass of the two prompt pions forming the $\rho^0$, $m_{\pip\pim} \in [0.6, 0.9] \gevcc$. This last requirement is the only selection step that enhances the $f_{CP}$ contribution to the $\Bz \to K_{\mathrm{res}} \gamma$ mode.

The $B_{\mathrm{tag}}$ vertex is reconstructed by combining the tracks in the event that have not been used in the reconstruction of the $B_{\mathrm{sig}}$~\cite{Belle-II:2024lwr}. The position of its decay vertex is determined by constraining the $B_{\mathrm{tag}}$ direction, computed as the difference between the position vector of the decay vertex and the position vector of the center of the interaction region, to be collinear with its momentum vector~\cite{Dey:2020dsr}.

We build a multivariate BDT classifier to distinguish true signal candidates from continuum events,
the main source of background. These continuum events feature a boosted jet-like topology whereas $\B\Bbar$ events result in a more isotropic distribution of final state particles. The different topologies are exploited in the BDT. The variables used for training the BDT are common between Belle and Belle~II, but each BDT is trained independently. The signal MC samples are used as a proxy for signal while the continuum MC samples are used as a proxy for background. We use a total of eleven variables in the BDT: the polar angle of the signal \B candidate in the c.m.\ frame, the angle between the $B_{\mathrm{sig}}$ thrust axis and the thrust axis of the rest of the event, and a total of nine modified Fox-Wolfram moments~\cite{PhysRevLett.41.1581,Belle:2003fgr}. We determine the optimal BDT output threshold by maximizing the signal significance, $N_S/\sqrt{N_S+N_B}$, where $N_S$ and $N_B$ denote the expected number of signal and background candidates, respectively, in a specific signal-enhanced window of $\de\in[-0.2,0.1]~\gev$ and $\mbc\in[5.27,5.29]~\gevcc$.
The BDT efficiency is about $65-70\%$ for signal and between $5-8\%$
for the continuum background.

There are multiple $B_{\mathrm{sig}}$ candidates in $15\%$ of events. As the final step of the candidate selection procedure, we select a single $B_{\mathrm{sig}}$ candidate in each event.
For Belle, we keep only the candidate with the best corrected $\chi^2$ of the vertex fit using a variable, $\xi$, discussed in Ref.~\cite{Tajima:2003bu}. The $\xi$ parameter is built specifically for time-dependent \CP measurements in Belle, because using the $\chi^2$ of the vertex fit directly would introduce a bias in the \dt distribution, since these are correlated. Unfortunately, the differences between the Belle and Belle~II vertexing process makes $\xi$ unusable for Belle~II, as it has been found to introduce a bias. Thus, for Belle~II, we randomly choose one candidate from the event. For events with multiple candidates these procedures select the correct candidate 45\% (65\%) of the time in Belle~II (Belle). We check using simulated samples that none of the selection criteria biases the measurement of the time-dependent \CP asymmetries. 

Using the simulated MC samples we can estimate the ratios of our final selection efficiencies for the different modes present in the $\Bz \to \KS \pip \pim \gamma$ decays:
\begin{equation}
    R_{K^*} = \frac{\epsilon_{K^*}}{\epsilon_{\KS \rho^0}} , \;    R_{(K\pi)_0} = \frac{\epsilon_{(K\pi)_0}}{\epsilon_{\KS \rho^0}},
\end{equation}
where $\epsilon_{K^*}$, $\epsilon_{(K\pi)_0}$ and $\epsilon_{\KS \rho^0}$ are the efficiencies of the decay modes quoted in Eq.~\ref{eq:kstarpath}, Eq.~\ref{eq:swavepath} and of the $f_{CP}\gamma$ decay, $\Bz \to \KS \rho^0\gamma$, respectively. These efficiency ratios are expected to be insensitive to potential mismodeling of the detector to first order, since these processes result in a common set of final state particles. Nevertheless, these efficiencies can be sensitive to the amplitude models considered for each decay mode, mainly because of the selection applied to the $m_{\pip\pim}$ distribution. Therefore, we also provide ratios of the efficiency without the $m_{\pip\pim}$ requirement. The estimated ratios are $R_{K^*} = 1.04\pm0.04\;(0.48\pm0.03)$ and $R_{(K\pi)_0}= 1.00\pm0.04\;(0.35\pm0.03)$ for the full selection without (with) the prompt pion pair mass,  $m_{\pip\pim}$, requirement. This information is not used in this analysis, but it is needed to make an interpretation of the measurement, i.e. to compute the proportion of the \CP mode with respect to the non-\CP modes and thus determine the value of the mixing-induced \CP observable $S_{\KS\rho^0\gamma}$.

The same procedure is applied to reconstruct and select our control mode ${B^+\to K^+ \pip \pim \gamma}$, which is also the isospin partner mode of $\Bz \to \KS \pip \pim \gamma$. We use it to validate different steps of the fit strategy. The candidate selection for this mode is performed in exactly the same manner as for the decay mode of interest, with the exception of requiring a $K^+$ instead of a \KS. The $K^+$ candidate must satisfy the same charged track requirements used for prompt pions, but must have PID information consistent with the kaon hypothesis. 
Since the \KS is not used to reconstruct the $B^0$ decay vertex, we do not use the $K^+$ track information to reconstruct the $B^+$ decay vertex.

\section{Time-dependent \CP fit strategy}
\label{fitstrat}

\subsection{Time-dependence and flavor tagger}

The \dt distribution is sensitive to the \CP parameters, $C$, $S$, \Sp and \Sm. In Belle, the \dt observable is computed as $\dt = \Delta z / \beta\gamma c$, where $\Delta z \equiv z_{\mathrm{sig}}-z_{\mathrm{tag}}$ is the difference between the $z$ positions of the $B_{\mathrm{sig}}$ and $B_{\mathrm{tag}}$ decay vertices and $\beta$ and $\gamma$ are the relativistic Lorentz factors. For Belle~II, instead of $\Delta z$, we make use of $\Delta l$, the distance of the $B_{\mathrm{sig}}$ and $B_{\mathrm{tag}}$ decay-vertex positions along the \ups boost direction and calculate $\dt = \Delta l / \beta\gamma c$. For Belle, $\beta\gamma \approx0.425$, while for Belle~II $\beta \gamma\approx0.284$, because the beam energy asymmetry is smaller. The precise value of $\beta\gamma$ is periodically calibrated using data. In Belle~II we apply a correction~\cite{Belle-II:2024hqw} for the small boost of the \B mesons in the c.m.\ frame~\cite{Bevan:2014iga}, to account for the transverse component of the $B$ flight in the laboratory frame. For Belle, the correction is not applied and the residual effect is taken into account as part of the resolution function~\cite{Tajima:2003bu}.

To ensure an accurate extraction of the time-dependent \CP asymmetry parameters, the physics probability density function (p.d.f.)\ is convolved with the finite \dt detector resolution, described by a resolution function $R(\dt)$. We use different parameterizations of the \dt resolution function for Belle and Belle~II.

The \dt resolution function for Belle~\cite{Tajima:2003bu} is the combination of four individual contributions, arising from the resolution on the $B_{\mathrm{sig}}$ decay vertex position ($z_{\mathrm{sig}}$), the resolution on the $B_{\mathrm{tag}}$ decay vertex position ($z_{\mathrm{tag}}$), the resolution effects induced by tertiary vertices, in particular due to tracks stemming from $D$ meson decays biasing the $B_{\mathrm{tag}}$ decay vertex, and lastly a contribution to correct for the small boost of the \B mesons in the c.m.\ frame. The resolution function is conditional on several per-candidate observables of the fitted signal and tag vertices: the number of tracks, the number of degrees of freedom, $\chi^2$ values and uncertainties on the longitudinal position of the vertices.

For Belle~II, we parameterize the \dt resolution function by combining several Gaussian-like distributions with a per-candidate dependence on the \dt uncertainty, $\sigma_{\dt}$. The description of the resolution function and its calibration is detailed in Ref.~\cite{Belle-II:2024lwr}.

The measurement of the time-dependent \CP asymmetries requires the flavor of the decaying signal \B meson.  Since the  $B_{\mathrm{tag}}$ and $B_{\mathrm{sig}}$ are entangled, by measuring the flavor of the tag \B when it decays we can infer the flavor of the signal \B. This premise is valid independently of which \B meson decays first, i.e.\ \dt can assume negative values.

The $B_{\mathrm{tag}}$ decay products are input into dedicated multi-variate analysis tools referred to as flavor-tagging (FT) algorithms. Different FT algorithms are used for the Belle and Belle~II measurements. We use the algorithm described in Ref.~\cite{Belle:2004uxp} for Belle, which achieves around a $30\%$ effective tagging efficiency. For Belle~II, we use a graph neural network to estimate the flavor of the $B_{\mathrm{tag}}$ meson, GFlat~\cite{Belle-II:2024lwr}. The GFlat algorithm provides an effective tagging efficiency of about 37\%.

The FT algorithm provides the flavor prediction $q$ ($q = 1$ for $B_{\mathrm{tag}}=\Bz$ and $q = - 1$ for $B_{\mathrm{tag}}=\Bzb$)
and the tag quality, $r$, which ranges from zero for no discriminating power to one for unambiguous flavor assignment. The imperfect assignment of the flavor by the FT algorithm is described by three parameters: the wrong-tag probability, $w$; the wrong-tag probability difference between \Bz and \Bzb, $\Delta w$; and the tagging efficiency asymmetry between \Bz and \Bzb, $a_{\mathrm{tag}}$.
The values of these parameters are obtained through data-driven calibration in seven bins of tag-quality $r\approx 1-2w$~\cite{Belle-II:2024lwr,Belle:2004uxp}. The $r$-calibration binning is defined as $[0,0.1,0.25,0.5,0.625,0.75,0.875,1]$ for Belle and as $[0,0.1,0.25,0.45,0.6,0.725,0.875,1]$ for Belle~II. For Belle, the calibration on the first bin $r\in[0,0.1]$ provides no inherent flavor discrimination (i.e.\ $w\equiv 0.5$) so this bin is not used for the measurement of the time-dependent \CP asymmetries. In both Belle and Belle~II, the output of the FT algorithm is used on a per-candidate basis when performing the fit. Additionally, for Belle~II, we split our sample into a good quality range $r \in [0, 0.875]$ and an excellent quality range $r \in [0.875, 1]$ and perform the fit simultaneously in these two $r$ regions. This provides improved statistical sensitivity by accounting for the different background levels.
For Belle, $a_{\mathrm{tag}}$ is neglected in the calibration procedure, whereas it is included in Belle~II.  The values of this parameter in both experiments are consistent with zero within uncertainties.

The p.d.f. describing the \dt distribution is:
\begin{align}
    P(\Delta t, q, w, & \Delta w, a_\mathrm{\mathrm{tag}}) = \frac{e^{-|\Delta t|/\tau_{\Bz}}}{4 \tau_{\Bz}} \big\{ 1 - q \Delta w + q a_\mathrm{\mathrm{tag}} (1 - 2w) \nonumber\\
    &~+ [q (1 - 2w) + a_\mathrm{\mathrm{tag}} (1 - q \Delta w)] [S \sin{(\Delta m \Delta t)} - C \cos{(\Delta m \Delta t)}] \big\}
    \otimes R(\Delta t),
    \label{eq:CPmodel_ft}
\end{align}
where $\tau_{\Bz}$ is the lifetime of \Bz meson, and $\otimes$ denotes the convolution of the physics p.d.f. with the resolution function, $R(\Delta t)$. 

\subsection{Extraction of the \CP parameters}

We measure the time-dependent \CP asymmetries by performing an unbinned maximum likelihood fit in $\mbc$, $\Delta E$ and $\dt$. The beam-energy constrained mass and energy difference, described previously, are powerful observables to disentangle the signal component from the background components. The \dt distribution is sensitive to the time-dependent \CP asymmetries as previously discussed. The fit is performed independently for Belle and Belle~II.

In addition to the signal component, three background sources are present in the fit region. The main background contribution is the one arising from continuum events: while these events are highly suppressed by the candidate selection, a significant number still remains. The second background contribution is from misreconstructed $\B\Bbar$ decays, including both \Bz and \Bp decays. These include several \B decays that can partially mimic our final state, mainly radiative \B decays to two hadrons (\KS, $\pi^\pm$) plus a particle from the $B_{\mathrm{tag}}$ side. The third background source is self cross-feed (SCF): true $\Bz \to \KS \pip \pim \gamma$ signal decays that are incorrectly reconstructed. The leading source of SCF is when a pion from the $B_{\mathrm{tag}}$ is reconstructed as one of the signal prompt pions.

We model each of the four components using the simulated MC samples, independently in each of the fit observables  ($\mbc, \Delta E, \dt$). Two exceptions are discussed later, affecting the \dt distribution of continuum candidates and the $\mbc$ and $\Delta E$ distributions of the contribution from misreconstructed $\B\Bbar$ candidates. The same modeling is used in the fit for Belle and Belle~II if not otherwise stated. The parameters for the modeling are fixed to those determined in fits to the simulated MC samples if not otherwise stated. The models for each observable and component are presented in the following.

For the signal component, the $\mbc$ and $\Delta E$ distributions are expected to peak at $\mbc= m_{\Bz}$ and $\Delta E =0$, where $m_{\Bz}$ is the mass of the \Bz meson. We describe both distributions with Crystal Ball functions~\cite{Gaiser:Phd,Skwarnicki:1986xj}, which combine a Gaussian core with power-law distributions for the tails. The mean and width parameters of the Gaussian cores are allowed to float freely in the fit to the data. The \dt fit function for the signal component was presented in Eq.~\ref{eq:CPmodel_ft}. The values of the \CP observables $S$ and $C$ are free parameters. The values of the \Bz lifetime and mass difference are fixed to the world-average values, $\tau_{\Bz} = 1.517\pm0.004\;\mathrm{ps}$  and $\Delta m = 0.5063 \pm 0.0019\;\mathrm{ps^{-1}}$~\cite{ParticleDataGroup:2024cfk} and their uncertainties are propagated as systematic uncertainties.

For the continuum component, we model the \mbc distribution using an ARGUS p.d.f. The threshold parameter of the ARGUS function is common to all contributions to the \mbc distribution and is free to float in the fit to the data. We model the $\de$ distribution with an exponential function. Different approaches are taken to model the \dt distribution in Belle and Belle~II. The Belle modeling uses the combination of a Gaussian distribution and an exponential convolved with a Gaussian~\cite{Tajima:2003bu}. We observe mismodeling between the \dt distribution in our continuum MC samples and the \dt distribution of the data samples below the \ups energy, where only continuum events are present. We correct our \dt distribution in the MC samples using weights obtained in 40 \dt bins from $-10\;\mathrm{ps}$ to $10\;\mathrm{ps}$. These weights are given by the ratio, in each \dt bin, of the yields of the data below the \ups energy to the continuum MC. We obtain the weights from the isospin partner mode.

For Belle~II, the \dt distribution is modeled by the sum of three Gaussian functions, describing the core, the tail, and the outliers of the resolution. The widths of the core and tail Gaussians are free to float in the fit to the data.

The misreconstructed $\B\Bbar$ component is modeled differently in all three distributions for Belle and Belle~II. In Belle, we model the \mbc distribution as the sum of a Gaussian and an ARGUS p.d.f. while we model the \de distribution with a sum of a Gaussian and an exponential distribution. The \dt distribution is modeled similarly to the continuum component, with a Gaussian and an exponential convolved with a Gaussian. It is worth noting that we find a similar quality of description if we instead use Eq.~\ref{eq:CPmodel_ft}, with both $C$ and $S$ fixed to zero; this alternative is used to evaluate the systematic uncertainty. For Belle~II, we describe the \mbc and \de distributions together using a kernel density estimator, to take into account small correlations between the two variables. The \dt distribution is modeled with Eq.~\ref{eq:CPmodel_ft} with both \CP observables, $C$ and $S$, fixed to zero. Most of the $\B\Bbar$ background arises from charged \B meson decays and other radiative decays, thus the values of the \CP parameters are expected to be zero or very close to it. The validity of this assumption is considered when assigning systematic uncertainties.

Lastly, for the SCF component, the $\mbc$ distribution is modeled by the sum of a bifurcated Gaussian distribution and an ARGUS p.d.f.~\cite{ARGUS:1990hfq}. We use a second order Chebychev polynomial to describe the $\de$ distribution. For Belle~II, we use a Chebychev polynomial with an additional bifurcated Gaussian distribution for improved modeling. The \dt distribution for the SCF component is also modeled with Eq.~\ref{eq:CPmodel_ft}, with $S_{\mathrm{SCF}} = \kappa_{\mathrm{SCF}}\cdot S$ and $C_{\mathrm{SCF}} = \kappa_{\mathrm{SCF}}\cdot C$,  where $\kappa_{\mathrm{SCF}}$ is common for both observables and is obtained from a fit to the simulated samples. In the unbinned maximum likelihood fit, a Gaussian constraint of mean $0.8$ and width $0.2$ is applied to $\kappa_{\mathrm{SCF}}$: this accounts for the statistical fluctuations of $\kappa_{\mathrm{SCF}}$ in the simulated samples.

\begin{figure*}[t]
\centering
    \includegraphics[width=0.47\textwidth]{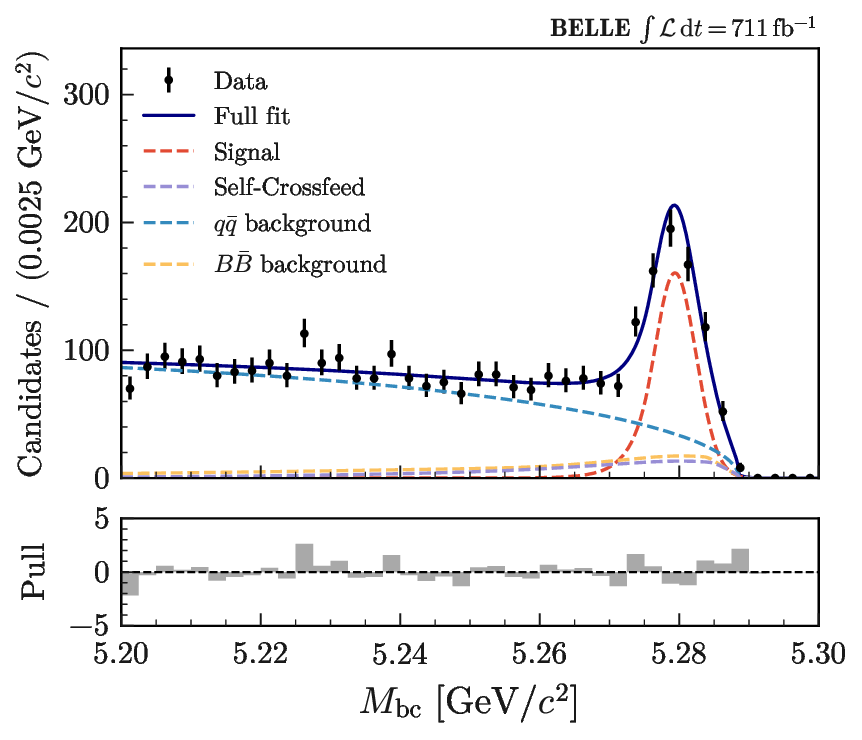}
    \hfill
    \includegraphics[width=0.47\textwidth]{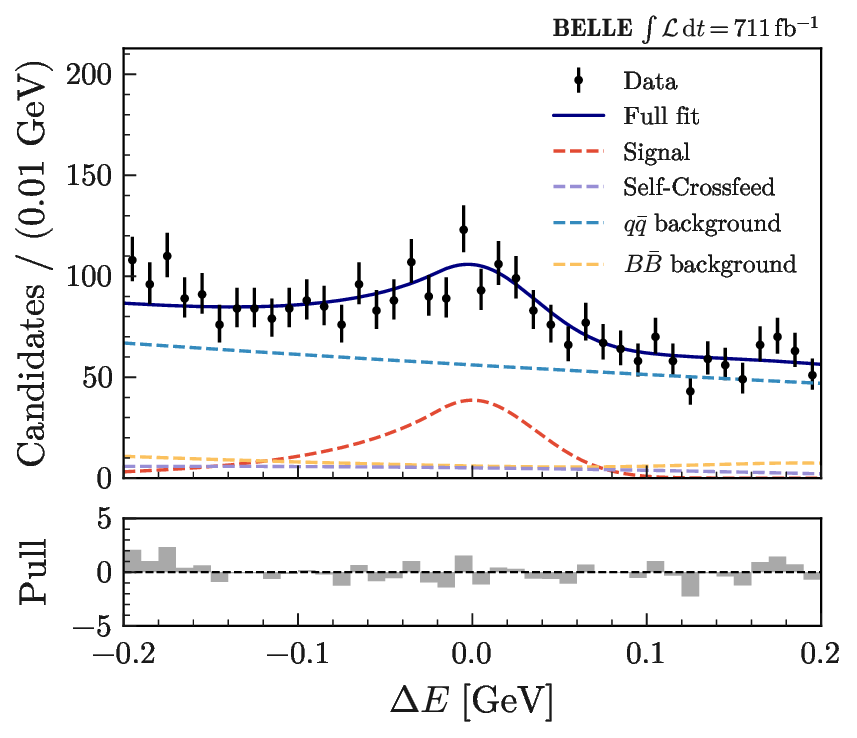}\\
    \includegraphics[width=0.47\textwidth]{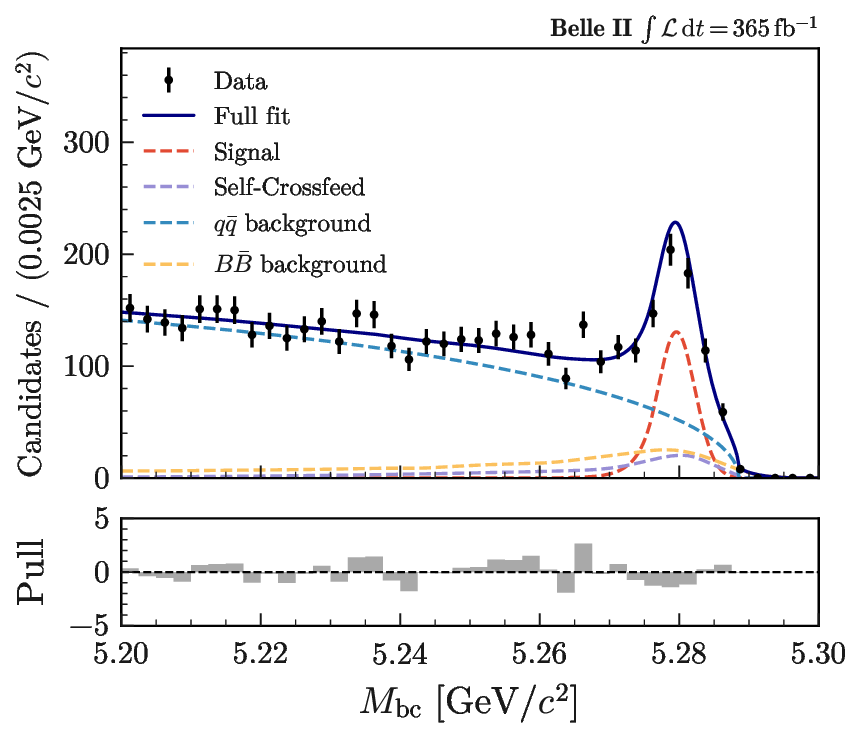}
    \hfill
    \includegraphics[width=0.47\textwidth]{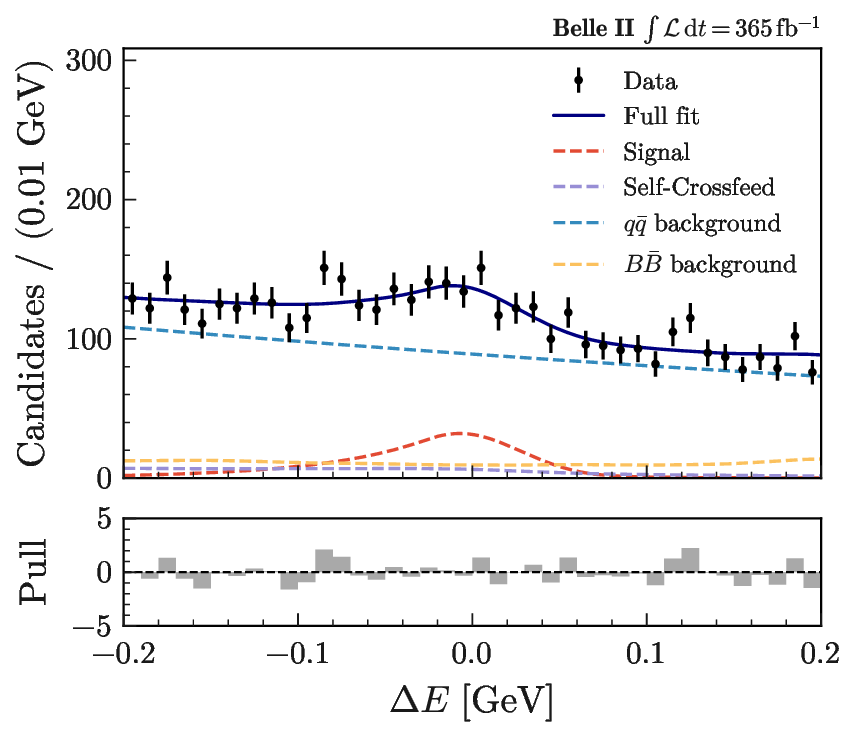}
    \caption{Unbinned maximum likelihood projections on \mbc (left) and \de (right) using the Belle (top) and Belle~II (bottom) datasets.}
    \label{fig:figure_1}
\end{figure*}

We allow three yields to be completely free in data when performing the unbinned maximum likelihood fit. The continuum component and the misreconstructed $\B\Bbar$ component have their yields free to float. The sum of the signal and SCF yields is free to float. The relative proportion of the signal and SCF components is fixed to the value extracted from the simulated MC samples. This value is around 29\% for Belle.  For Belle~II it is 38\% for the first $r$-region and $27\%$ for the second $r$-region.

Additionally, for Belle~II, as previously described, a simultaneous fit is performed in two regions of flavor tagging quality, $r$. The yields in each $r$ bin are free to float. Consequently, for Belle we have a total of 10 free parameters: $S$, $C$, three yields, and five modeling parameters. In Belle~II, we have 15 free parameters, due to the additional two free parameters in the \dt distribution of the continuum component and the three additional yields, due to the additional $r$-region.

\begin{figure*}[t]
\centering
    \includegraphics[width=0.47\textwidth]{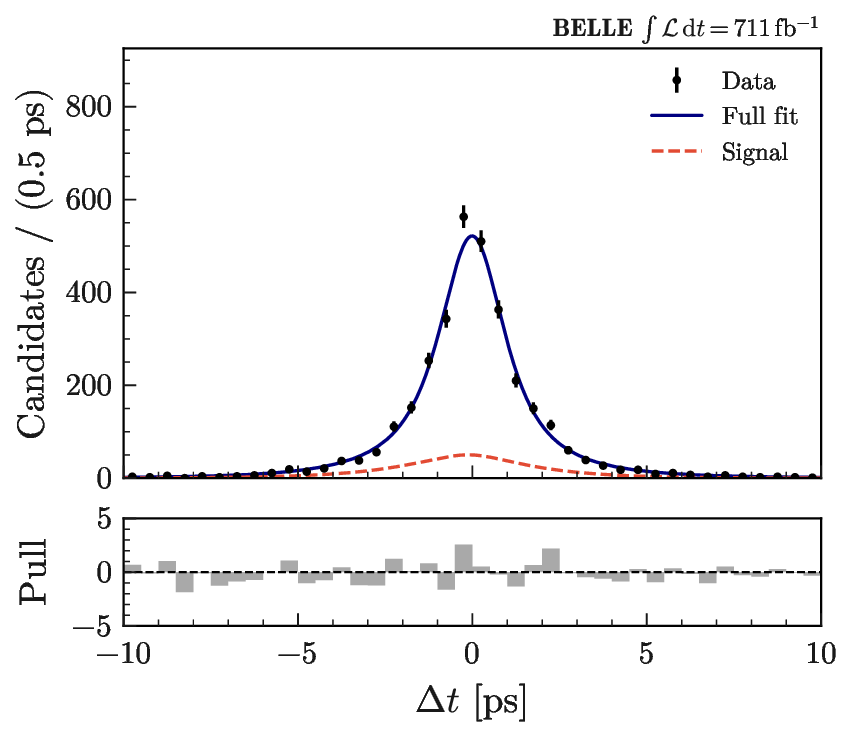}
    \hfill
    \includegraphics[width=0.47\textwidth]{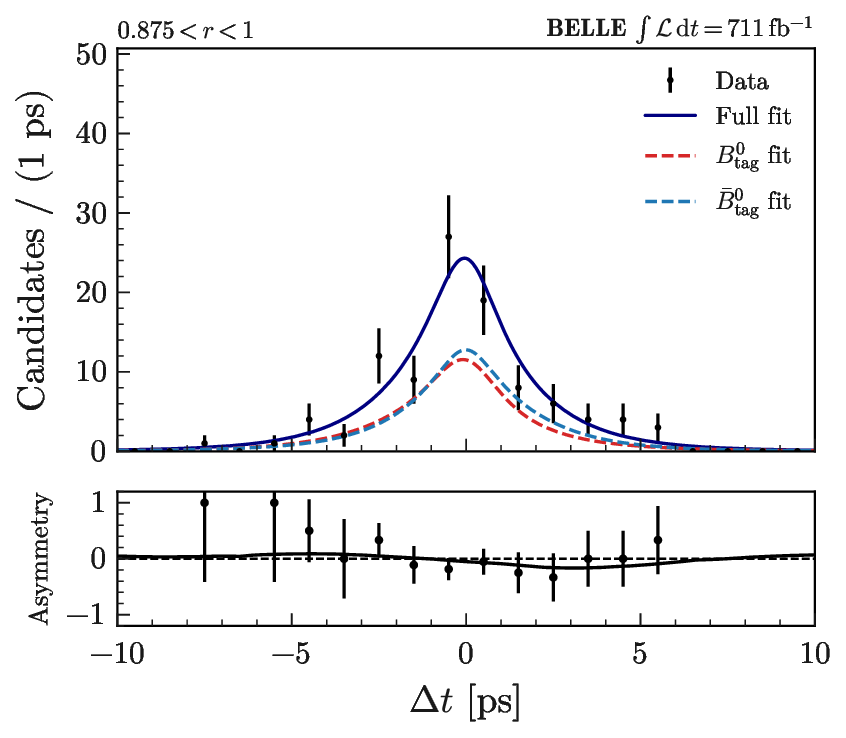}\\
    \includegraphics[width=0.47\textwidth]{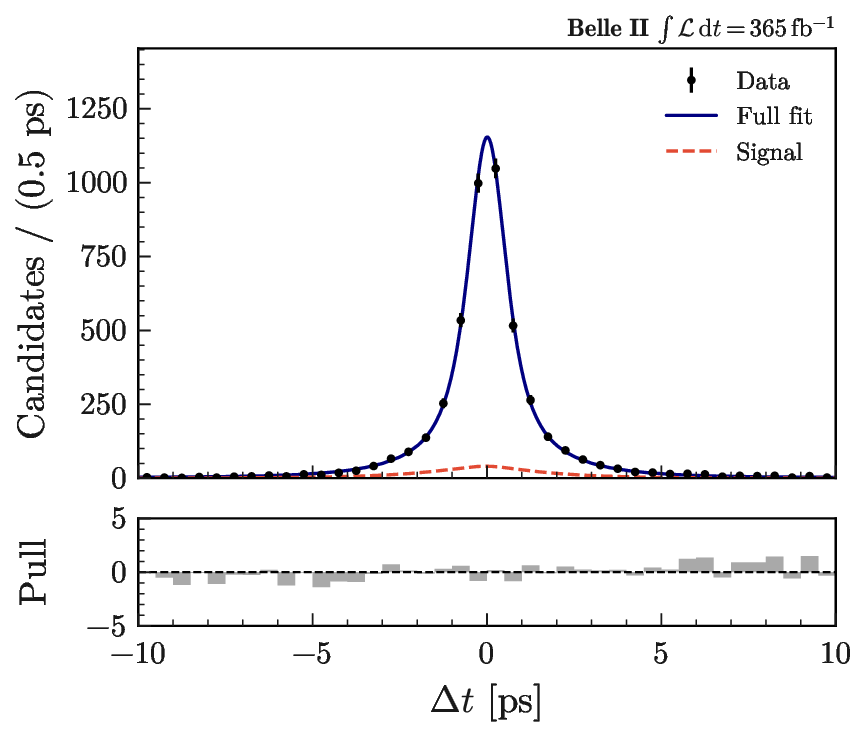}
    \hfill
    \includegraphics[width=0.47\textwidth]{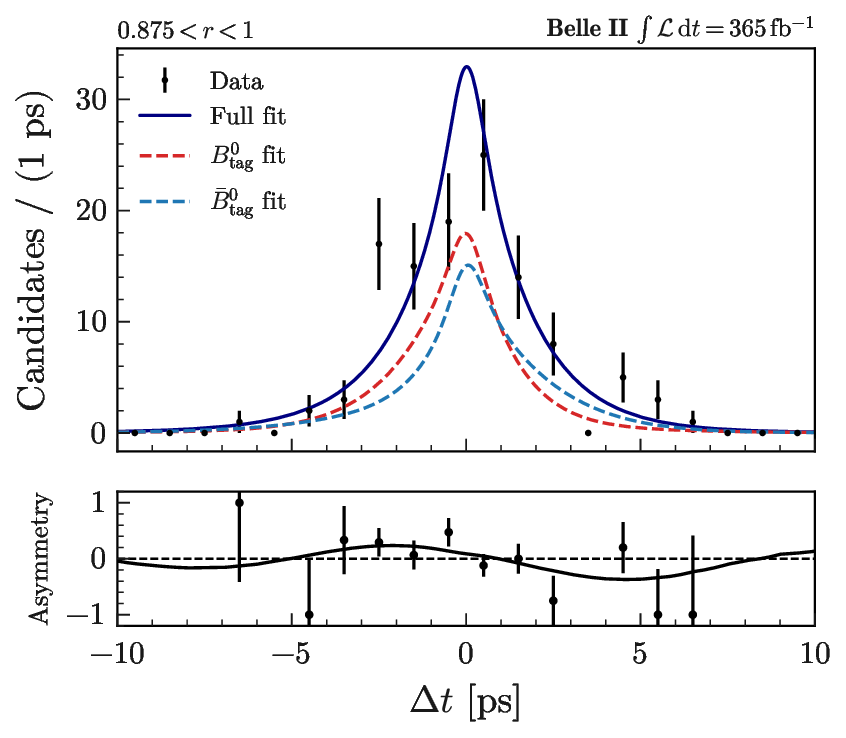}
    \caption{Unbinned maximum likelihood projection on \dt (left) and unbinned maximum likelihood projection on \dt split by tagged \Bz and \Bzb candidates in a signal enhanced window defined by: $\mbc>5.27$~\gevcc, $|\de|< 0.1$~\gev and $r \in [0.875,1]$ (right), using the Belle (top) and Belle~II (bottom) dataset.}
    \label{fig:figure_2}
\end{figure*}

The results of unbinned maximum likelihood fits to \mbc, \de and \dt in the signal channel
are $C=-0.04\pm 0.11$ and $S=-0.18 \pm 0.17$, for the Belle dataset, and $C=-0.29\pm 0.13$ and $S= -0.36 \pm 0.16$, for the Belle~II dataset, where the uncertainties are statistical only. The correlations between the \CP parameters are $+4.8\%$ for Belle and $+10.4\%$ for Belle~II. We obtain $475\pm 31$ signal candidates for Belle and $350\pm 23$ for Belle~II. The projections of the fit result in \mbc and \de are shown in Fig.~\ref{fig:figure_1} for Belle and Belle~II. The projections on \dt are shown in Fig.~\ref{fig:figure_2}. The \dt projections for \Bz- and \Bzb-tagged events are also shown, together with the asymmetry, in a signal enhanced region. The signal over background ratio in this region is about 3.4 for Belle and 2.4 for Belle~II.

The same fit strategy is used to extract the \CP observables using the two halves of the $({m^2(\KS \pip), m^2(\KS \pim))}$  plane, \Sp and \Sm. Candidates in the two halves of the plane are fit simultaneously. All free-floating parameters are common in the simultaneous fit to the two halves with the exception of the yields. This unbinned maximum likelihood fit is independent of the previous one. It has a total of 14 free floating parameters for Belle, due to an additional \CP observable: $C$, \Sp and \Sm  (instead of only $C$ and $S$) and a total of six free floating yields instead of three. For Belle~II the number of free-floating parameters rises to 22.

The results of the simultaneous unbinned maximum likelihood fit in the two half-planes are $\Sp=-0.33\pm0.34$ and $\Sm=-0.36\pm0.38$ for Belle and $\Sp=-0.72\pm0.31$ and $\Sm=0.70\pm0.30$ for Belle~II, with correlations of $-5.9\%$ and $+6.5\%$, respectively.

\subsection{Validation and systematic uncertainties}
\label{systs}
We validate the fit strategy using the simulated MC samples for both the signal mode $\Bz \to \KS \pip \pim \gamma$ and the control mode ${B^+\to K^+ \pip \pim \gamma}$. Then we validate on the data using the control mode and finally we perform the time-dependent \CP asymmetry measurement.

The strategy for extracting the \CP observables is first validated with MC samples. Using our nominal model, we create and fit a large number of data-like MC samples, with a luminosity equivalent to that used in the measurement, by resampling with replacement the combined MC samples of all components. The signal component in these data-like samples is generated with one of several different values of the mixing-induced \CP observable, $S$. The differences between the fitted and expected values of the \CP observables (and yields) from this validation process are taken as systematic uncertainties.

We perform the measurement of the time-dependent \CP asymmetries in the control mode using the same procedure as for the signal mode, and the same values for the $\Delta m$ and FT parameters. No time-dependent \CP asymmetry is expected in \Bp decays; we find results for $C$ and $S$ compatible with zero within one standard deviation. Additionally, we fit for the \Bp lifetime and obtain, $\tau_{\B^+}=1.62\pm0.06\;(1.71\pm0.06)\;\mathrm{ps}$ for Belle~II (Belle), which are compatible with the world average value within one standard deviation.

As a last validation step, we fit the \Bz lifetime in the data samples for $\Bz \to \KS \pip \pim \gamma$ signal candidates, obtaining $\tau_{\Bz} = 1.52\;(1.47)\pm 0.11\;(0.10)\;\mathrm{ps}$ for Belle~II (Belle), compatible with the world-average value.

We consider several sources of systematic uncertainty that may affect the time-dependent \CP asymmetry measurement. These are computed independently for the Belle and Belle~II datasets. We list the sources in Table~\ref{tab:table_1} for Belle and Table~\ref{tab:table_2} for Belle~II. The majority of the uncertainties are evaluated by fitting the samples with alternative fit models and comparing the results to those from the nominal fit. The main systematic uncertainties arise from the vertex detector misalignment for Belle and the differences of the \CP observables obtained as part of the validation process for Belle~II.

We evaluate the systematic uncertainties arising from the parameterization of the fit distributions, the FT calibration, the resolution function model, and the usage of external parameters $(\tau_{\Bz}, \Delta m)$, in a similar manner. These are evaluated by fitting the data samples with alternative values of the corresponding fixed parameters. The alternative values are obtained by drawing values of each parameter from a normal distribution whose width is set by the parameter uncertainty or the covariance matrix obtained from MC simulated samples. The systematic uncertainty is computed as the standard deviation of the distribution of the fitted \CP observables that are obtained from the alternative fits.

Additionally, we evaluate a systematic uncertainty due to the fixed ratio between the signal and the SCF by varying it by $\pm20\%$ and refitting. This variation accounts for the statistical fluctuations of this ratio in the simulated samples. The shift in the \CP observables is assigned as the systematic uncertainty.

We use the validation process to assign a systematic uncertainty due to possible biases on the extracted yields and \CP observables.  For the yields, we perform an alternative fit to the validation MC samples with the yields fixed to the generated values, and take the difference of the resulting \CP observable values with respect to the nominal fit as the associated uncertainty.
For the \CP observables, we fit the validation MC samples using the nominal model and take the average deviation from the generated \CP observables, for different values of $S$, as the systematic uncertainty. This uncertainty accounts for various contributing factors, including the limited size of the MC samples used in the validation as well as any discrepancies among the \mbc, \de and \dt distributions in the MC samples and the models used to describe them. Although this systematic uncertainty is small compared to the statistical uncertainty, it remains the dominant contribution to the overall systematic uncertainty for Belle~II.

A systematic uncertainty is computed for the tag-side interference. This effect arises from neglecting the interference between the different amplitudes present in the final state of the $B_{\mathrm{tag}}$ decay mode (tag-side). These interference effects can directly alter the measurement of the \CP observables as described in Ref.~\cite{Long:2003wq}. The corresponding uncertainty is the only one that is correlated between Belle and Belle~II.

A systematic uncertainty is computed for possible \CP violation in the misreconstructed $\B\Bbar$ background component. The values of the assumed \CP observables for the misreconstructed $\B\Bbar$ background are varied in the data samples: each of $C_{\B\Bbar}$ and $ S_{\B\Bbar}$ is independently assigned a value of $\pm 0.1$. The maximum difference of these four combinations relative to the nominal fit is assigned as the systematic uncertainty. The magnitude of this uncertainty is different between Belle and Belle~II. The nominal \dt model for the misreconstructed $\B\Bbar$ background component for Belle, as previously presented, is not Eq.~\ref{eq:CPmodel_ft}, whereas to compute this systematic we use Eq.~\ref{eq:CPmodel_ft} as the alternative model.

Lastly, we evaluate systematic uncertainties arising from a residual vertex detector misalignment: we reconstruct the \B candidates with various misalignment scenarios using simulated MC samples, leading to slightly different fit results. These variations are performed separately for Belle and Belle~II, given the intrinsic detector differences. In particular, for Belle, this irreducible source of systematic uncertainty is obtained for $S$ and $C$, and the uncertainties for \Sp and \Sm are computed from $S$ as fully correlated. This uncertainty dominates the overall systematic uncertainty for Belle.
\begin{table*}[ht]
    \centering
    \caption{Systematic uncertainties for the Belle measurement.}
        \begin{tabular}{c|cccc}
        \hline
        {\bf Source of uncertainty}  &  $C$  &  $S$  &  $\Sp$  &  $\Sm$  \\ 
        \hline
        \hline
        Fixed shape parameters  &  $0.003$  &  $0.002$  &  $0.004$  &  $0.004$  \\
        Flavor Tagging parameters  &  $0.003$  &  $0.002$  &  $0.006$  &  $0.006$  \\
        Resolution function parameters  &  $0.009$  &  $0.033$  &  $0.063$  &  $0.027$  \\
        $\tau_{\Bz}$ \& $\Delta m$  &  $0.001$  &  $0.001$  &  $0.002$  &  $0.002$  \\
        Fixed SCF fraction  &  $0.005$  &  $0.006$  &  $0.010$  &  $<0.001$  \\
        Yield bias  &  $0.001$  &  $0.008$  &  $0.015$  &  $0.002$  \\
        CP fit validation  &  $0.018$  &  $0.019$  &  $0.035$  &  $0.075$  \\
        Tag-side interference  &  $0.028$  &  $<0.001$  &  $<0.001$  &  $<0.001$  \\
        CP violation in $\B\Bbar$ background  &  $0.047$  &  $0.039$  &  $0.060$  &  $0.079$  \\
        Residual misalignment  &  $0.03$  &  $0.06$  &  $0.12$  &  $<0.001$ \\
       \hline
        Total systematic uncertainty  &  $0.066$  &  $0.081$  &  $0.153$  &  $0.112$  \\
        \hline
        \end{tabular}
    \label{tab:table_1}
\end{table*}

\begin{table*}[ht]
    \centering
    \caption{Systematic uncertainties for the Belle~II measurement.}
    \begin{tabular}{c|cccc}
        \hline
        {\bf Source of uncertainty}  &  $C$  &  $S$  &  $\Sp$  &  $\Sm$  \\ 
        \hline
        \hline
        Fixed shape parameters  &  $0.003$  &  $0.005$  &  $0.005$  &  $0.004$  \\
        Flavor Tagging parameters  &  $0.018$  &  $0.007$  &  $0.014$  &  $0.012$  \\
        Resolution function parameters  &  $0.005$  &  $0.014$  &  $0.023$  &  $0.018$  \\
        $\tau_{\Bz}$ \& $\Delta m$  &  $<0.001$  &  $0.001$  &  $0.001$  &  $0.003$  \\
        Fixed SCF fraction  &  $0.006$  &  $0.004$  &  $0.008$  &  $0.011$  \\
        Yield bias  &  $0.005$  &  $0.004$  &  $0.008$  &  $0.014$  \\
        CP fit validation  &  $0.027$  &  $0.054$  &  $0.117$  &  $0.033$  \\
        Tag-side interference  &  $0.028$  &  $<0.001$  &  $<0.001$  &  $<0.001$  \\
        CP violation in $\B\Bbar$ background  &  $0.019$  &  $0.017$  &  $0.034$  &  $0.001$  \\
        Residual misalignment  &  $0.005$  &  $0.003$  &  $0.006$  &  $0.012$   \\
        \hline
        Total systematic uncertainty  &  $0.048$  &  $0.059$  &  $0.126$  &  $0.045$  \\
        \hline
    \end{tabular}
    \label{tab:table_2}
\end{table*}

\section{Results and conclusions}
\label{results}

The measured values of the \CP parameters in the decay $\Bz \to \KS \pip \pim \gamma$ from the Belle data are: 
\begin{align} \label{results_belle_final}
\centering
C \phantom{^{+}} &= -0.04 \pm 0.11 \pm 0.07 \notag \\
S  \phantom{^{+}} &= -0.18 \pm 0.17 \pm 0.08
\end{align}
and for the \CP measurement using the half-planes:
\begin{align} \label{results_belle_final_D}
\Sp  &= -0.33 \pm 0.34 \pm 0.15 \notag \\
\Sm  &= -0.36 \pm 0.38\pm 0.11
\end{align}
Using the Belle~II data we measure:
\begin{align} \label{results_belle2_final}
\centering
C \phantom{^{+}}  &= -0.29 \pm 0.13 \pm 0.05 \notag\\
S  \phantom{^{+}} &= -0.36 \pm 0.16 \pm 0.06
\end{align}
and for the \CP measurement using the half-planes:
\begin{align} \label{results_belle2_final_D}
\Sp  &= -0.72 \pm 0.31\pm 0.13 \notag\\
\Sm  &= \phantom{-}0.70 \pm 0.30\pm 0.05
\end{align}
where the second term corresponds to the statistical uncertainty and the third term corresponds to the systematic uncertainty. It is worth pointing out that since we compute the \CP parameters $S$ and $\{\Sp, \Sm\}$ using the same datasets, their values are fully correlated.

The measurements of $S$, $C$ and \Sp are in agreement for the two experiments, and compatible with the SM prediction within $1\sigma$ for Belle, and $2.3\sigma$ for Belle~II. The values obtained for \Sm are in slight tension ($2.2\sigma$ apart), but are independently compatible with the SM prediction of zero.

We combine the results following the method presented in Ref.~\cite{HeavyFlavorAveragingGroupHFLAV:2024ctg}, taking into account the correlations between the measured \CP parameters. We consider all sources of systematic uncertainty as uncorrelated, except the tag-side interference, which is fully correlated. 
The combined results are:

\begin{align}
C \phantom{^{+}} &= -0.17 \pm 0.09 \pm 0.04 \notag \\
S         \phantom{^{+}}   &= -0.29 \pm 0.11 \pm 0.05
\end{align}
and 
\begin{align}
S^{+}                   &= -0.57 \pm 0.23 \pm 0.10 \notag \\
S^{-}                   &= \phantom{-}0.31 \pm 0.24 \pm 0.05
\label{results_comb}
\end{align}
The correlation between $C$ and $S$ is $-3\%$, while the correlation between \Sp and \Sm is $+2\%$.

The value of $S$ measured in this paper can be transformed into $S_{\KS\rho^0\gamma}$ if combined with a dilution factor, which takes into account the different proportions of the \CP and non-\CP eigenstates. This last observable provides direct constraints to the Wilson coefficients, $C_7$ and $C_7'$.

The values of $\{\Sp, \Sm\}$ can be combined with two dilution-like factors, $\{a, b\}$, described in Ref.~\cite{Akar:2018zhv}, to provide different constraints to the same Wilson coefficients.

Our combined measurement of $S$ and $C$ improves by at least a factor of two the corresponding uncertainties on the \CP observables measured previously by the Belle and BaBar collaborations. This result supersedes the time-dependent \CP asymmetry, namely $\mathcal{S_{\mathrm{eff}}}$, presented in Ref.~\cite{Belle:2008fjm}. We also measure, for the first time, the \CP parameters \Sp and \Sm, which are needed to apply additional constraints to new physics models with enhanced right-handed currents that may affect the ${\Bz \to \KS \pip \pim \gamma}$ transition.

This work, based on data collected using the Belle II detector, which was built and commissioned prior to March 2019,
and data collected using the Belle detector, which was operated until June 2010,
was supported by
Higher Education and Science Committee of the Republic of Armenia Grant No.~23LCG-1C011;
Australian Research Council and Research Grants
No.~DP200101792, 
No.~DP210101900, 
No.~DP210102831, 
No.~DE220100462, 
No.~LE210100098, 
and
No.~LE230100085; 
Austrian Federal Ministry of Education, Science and Research,
Austrian Science Fund (FWF) Grants
DOI:~10.55776/P34529,
DOI:~10.55776/J4731,
DOI:~10.55776/J4625,
DOI:~10.55776/M3153,
and
DOI:~10.55776/PAT1836324,
and
Horizon 2020 ERC Starting Grant No.~947006 ``InterLeptons'';
Natural Sciences and Engineering Research Council of Canada, Digital Research Alliance of Canada, and Canada Foundation for Innovation;
National Key R\&D Program of China under Contract No.~2024YFA1610503,
and
No.~2024YFA1610504
National Natural Science Foundation of China and Research Grants
No.~11575017,
No.~11761141009,
No.~11705209,
No.~11975076,
No.~12135005,
No.~12150004,
No.~12161141008,
No.~12405099,
No.~12475093,
and
No.~12175041,
and Shandong Provincial Natural Science Foundation Project~ZR2022JQ02;
the Czech Science Foundation Grant No. 22-18469S,  Regional funds of EU/MEYS: OPJAK
FORTE CZ.02.01.01/00/22\_008/0004632 
and
Charles University Grant Agency project No. 246122;
European Research Council, Seventh Framework PIEF-GA-2013-622527,
Horizon 2020 ERC-Advanced Grants No.~267104 and No.~884719,
Horizon 2020 ERC-Consolidator Grant No.~819127,
Horizon 2020 Marie Sklodowska-Curie Grant Agreement No.~700525 ``NIOBE''
and
No.~101026516,
and
Horizon 2020 Marie Sklodowska-Curie RISE project JENNIFER2 Grant Agreement No.~822070 (European grants);
L'Institut National de Physique Nucl\'{e}aire et de Physique des Particules (IN2P3) du CNRS
and
L'Agence Nationale de la Recherche (ANR) under Grant No.~ANR-23-CE31-0018 (France);
BMFTR, DFG, HGF, MPG, and AvH Foundation (Germany);
Department of Atomic Energy under Project Identification No.~RTI 4002,
Department of Science and Technology,
and
UPES SEED funding programs
No.~UPES/R\&D-SEED-INFRA/17052023/01 and
No.~UPES/R\&D-SOE/20062022/06 (India);
Israel Science Foundation Grant No.~2476/17,
U.S.-Israel Binational Science Foundation Grant No.~2016113, and
Israel Ministry of Science Grant No.~3-16543;
Istituto Nazionale di Fisica Nucleare and the Research Grants BELLE2,
and
the ICSC – Centro Nazionale di Ricerca in High Performance Computing, Big Data and Quantum Computing, funded by European Union – NextGenerationEU;
Japan Society for the Promotion of Science, Grant-in-Aid for Scientific Research Grants
No.~16H03968,
No.~16H03993,
No.~16H06492,
No.~16K05323,
No.~17H01133,
No.~17H05405,
No.~18K03621,
No.~18H03710,
No.~18H05226,
No.~19H00682, 
No.~20H05850,
No.~20H05858,
No.~22H00144,
No.~22K14056,
No.~22K21347,
No.~23H05433,
No.~26220706,
and
No.~26400255,
and
the Ministry of Education, Culture, Sports, Science, and Technology (MEXT) of Japan;  
National Research Foundation (NRF) of Korea Grants
No.~2021R1-F1A-1064008, 
No.~2022R1-A2C-1003993,
No.~2022R1-A2C-1092335,
No.~RS-2016-NR017151,
No.~RS-2018-NR031074,
No.~RS-2021-NR060129,
No.~RS-2023-00208693,
No.~RS-2024-00354342
and
No.~RS-2025-02219521,
Radiation Science Research Institute,
Foreign Large-Size Research Facility Application Supporting project,
the Global Science Experimental Data Hub Center, the Korea Institute of Science and
Technology Information (K25L2M2C3 ) 
and
KREONET/GLORIAD;
Universiti Malaya RU grant, Akademi Sains Malaysia, and Ministry of Education Malaysia;
Frontiers of Science Program Contracts
No.~FOINS-296,
No.~CB-221329,
No.~CB-236394,
No.~CB-254409,
and
No.~CB-180023, and SEP-CINVESTAV Research Grant No.~237 (Mexico);
the Polish Ministry of Science and Higher Education and the National Science Center;
the Ministry of Science and Higher Education of the Russian Federation
and
the HSE University Basic Research Program, Moscow;
University of Tabuk Research Grants
No.~S-0256-1438 and No.~S-0280-1439 (Saudi Arabia), and
Researchers Supporting Project number (RSPD2025R873), King Saud University, Riyadh,
Saudi Arabia;
Slovenian Research Agency and Research Grants
No.~J1-50010
and
No.~P1-0135;
Ikerbasque, Basque Foundation for Science,
State Agency for Research of the Spanish Ministry of Science and Innovation through Grant No. PID2022-136510NB-C33, Spain,
Agencia Estatal de Investigacion, Spain
Grant No.~RYC2020-029875-I
and
Generalitat Valenciana, Spain
Grant No.~CIDEGENT/2018/020;
the Swiss National Science Foundation;
The Knut and Alice Wallenberg Foundation (Sweden), Contracts No.~2021.0174, No.~2021.0299, and No.~2023.0315;
National Science and Technology Council,
and
Ministry of Education (Taiwan);
Thailand Center of Excellence in Physics;
TUBITAK ULAKBIM (Turkey);
National Research Foundation of Ukraine, Project No.~2020.02/0257,
and
Ministry of Education and Science of Ukraine;
the U.S. National Science Foundation and Research Grants
No.~PHY-1913789 
and
No.~PHY-2111604, 
and the U.S. Department of Energy and Research Awards
No.~DE-AC06-76RLO1830, 
No.~DE-SC0007983, 
No.~DE-SC0009824, 
No.~DE-SC0009973, 
No.~DE-SC0010007, 
No.~DE-SC0010073, 
No.~DE-SC0010118, 
No.~DE-SC0010504, 
No.~DE-SC0011784, 
No.~DE-SC0012704, 
No.~DE-SC0019230, 
No.~DE-SC0021274, 
No.~DE-SC0021616, 
No.~DE-SC0022350, 
No.~DE-SC0023470; 
and
the Vietnam Academy of Science and Technology (VAST) under Grants
No.~NVCC.05.02/25-25
and
No.~DL0000.05/26-27.

These acknowledgements are not to be interpreted as an endorsement of any statement made
by any of our institutes, funding agencies, governments, or their representatives.

We thank the SuperKEKB team for delivering high-luminosity collisions;
the KEK cryogenics group for the efficient operation of the detector solenoid magnet and IBBelle on site;
the KEK Computer Research Center for on-site computing support; the NII for SINET6 network support;
and the raw-data centers hosted by BNL, DESY, GridKa, IN2P3, INFN, 
PNNL/EMSL, 
and the University of Victoria.

\bibliographystyle{JHEP}
\bibliography{references}

\end{document}